\documentclass{amia}
\usepackage{graphicx}
\usepackage[labelfont=bf]{caption}
\usepackage{amsmath}
\usepackage[superscript,nomove]{cite}
\usepackage{color}
\usepackage{xcolor}
\usepackage{soul}
\begin{document}

\title{More Generalizable Models For Sepsis Detection Under Covariate Shift}

\author{Jifan Gao, MS$^{1}$, Philip L. Mar, MD$^{2}$, Guanhua Chen, PhD$^{1}$}

\institutes{
    $^1$University of Wisconsin, School of Medicine and Public Health\\
    $^2$Saint Louis University, School of Medicine \\
}

\maketitle

\noindent{\bf Abstract}

\textit{Sepsis is a major cause of mortality in the intensive care units (ICUs). Early intervention of sepsis can improve clinical outcomes for sepsis patients\cite{sakr2018sepsis,rhee2016diagnosing,singer2016third}. Machine learning models have been developed for clinical recognition of sepsis\cite{mitra2018sepsis,mao2018multicentre,pollard2018eicu}. A common assumption of supervised machine learning models is that the covariates in the testing data follow the same distributions as those in the training data. When this assumption is violated (e.g., there is covariate shift), models that performed well for training data could perform badly for testing data. Covariate shift happens when the relationships between covariates and the outcome stay the same, but the marginal distributions of the covariates differ among training and testing data. Covariate shift could make clinical risk prediction model nongeneralizable. In this study, we applied covariate shift corrections onto common machine learning models and have observed that these corrections can help the models be more generalizable under the occurrence of covariate shift when detecting the onset of sepsis.}

\section*{Introduction}
Sepsis is a life-threatening complication of infection, which can cause a cascade of changes that damage multiple organs and sometimes even leads to death\cite{sakr2018sepsis}. About 6 million people die from complications of sepsis each year\cite{kumar2006duration}. Early intervention with fluid resuscitation and antibiotics greatly improves the chance of survival for sepsis patients. However, the detection of sepsis is challenging because sepsis is a very heterogeneous syndrome\cite{mitra2018sepsis}. Rule-based scoring systems have been widely used in hospitals for identifying sepsis\cite{subbe2006validation,rangel1995natural,vincent1996sofa}. In recent years, with the prevalence of electronic health records (EHRs), many attempts have been made to build machine learning models for sepsis detection, and some of them outperform rule-based models. Lyra et al.\cite{lyra2019early} used 40 features for early prediction of sepsis using random forests. Mitra et al.\cite{mitra2018sepsis} and Mao et al.\cite{mao2018multicentre} studied several machine learning models using only six vital signs, including heart rate, respiratory rate, SpO2, temperature, systolic blood pressure, and diastolic blood pressure, and found machine learning models outperform rule-based scoring systems in sepsis detection task for ICU patients.

Generalizability refers to the ability of machine learning models to make correct predictions on  data collected from a different source that is not included in the training data \cite{dexter2020generalization}. A generalizable model should perform well for both training data and testing data; however, when there is data shift, it is difficult to ensure generalizability for machine learning models. Data shift is defined as when the population characteristics on which the model was developed is different from the population characteristics on which the model is applied\cite{quionero2009dataset, moreno2012unifying}. There are three types of data shift: covariate shift, prior probability shift, and concept shift. Covariate shift is associated with the change of distributions of the predictors; prior probability shift is related to the change of the outcomes; concept shift refers to the change of the underlying relationship between the predictors and the outcomes. Studies have shown that data shift can hurt the generalizability of machine learning models. For example, Hwang et al.\cite{hwang2019development} trained a model to detect abnormal chest radiographs and the model's specificity at a fixed threshold varied from 0.566 to 1.000 when validated using external data from different sites. Nestor et al. observed an AUROC drop of 0.29 for mortality prediction when models were trained on historical data and tested on future data. In this study, we focused on mitigating the model performance deterioration caused by covariate shift. 

In common machine learning models, there is presumption that the joint distribution of covariates/predictors and the outcome variable is the same in training and testing sets.
However, this assumption is violated when there is covariate shift, which refers to the situations where the underlying relationship between covariates and outcomes stays the same but training and testing sets follow different covariates distributions\cite{quionero2009dataset}. In clinical care scenarios, covariate shift is likely to occur due to temporal or geographical differences in populations. Some research has been done to demonstrate the need of covariate shift correction on clinical risk prediction tasks such as mortality and readmission\cite{nestor2019feature,curth2019transferring}. When the covariate information of the testing data is also available, a general framework called importance sampling (IS) weights has been proposed to remedy the impact of covariate shift. In particular, we would assign different weights to training samples based on their similarity to the test samples \cite{sugiyama2007covariate}. Higher weights are assigned to those samples which are common in the test set, and lower weights are assigned to those which rarely occur in the test set. The optimal sample weight for a subject in the training set with its covariates vector equals to $\mathbf{x}$ can be written as a density ratio as:
\begin{equation}
    r(\mathbf{x}) = \frac{p_{test}(\mathbf{x})}{p_{train}(\mathbf{x})}
\end{equation}
where $p_{test}(\mathbf{x})$ and $p_{train}(\mathbf{x})$ are the densities of covariates $\mathbf{x}$ associated with the testing and training set. To see why density ratio is useful, we show that when the covariate shift takes place, the expected (squared) prediction error of a model on its testing set can be written as:
\begin{equation}\label{eq2}
\begin{split}
    E_{test}(y-f(\mathbf{x}))^2 
    & = \int \int (y-f(\mathbf{x}))^2 p_{test}(\mathbf{x})p_{test}(y|\mathbf{x}) d\mathbf{x} \\
    & = \int \int (y-f(\mathbf{x}))^2 r(\mathbf{x}) p_{train}(\mathbf{x})p_{test}(y|\mathbf{x}) d\mathbf{x} \\
    & = \int \int (y-f(\mathbf{x}))^2 r(\mathbf{x}) p_{train}(\mathbf{x})p_{train}(y|\mathbf{x}) d\mathbf{x} \\
    & = E_{train}[ r(\mathbf{x}) (y-f(\mathbf{x}))^2 ] 
\end{split}
\end{equation}

Note that second equality holds since the conditional probability $p_{train}(y | \mathbf{x})$ and $p_{test}(y | \mathbf{x})$ are assumed to be equal under covariate shift. 
The quantity $E_{train}[ r(\mathbf{x}) (y-f(\mathbf{x}))^2 ]$ can be empirically estimated given the training data and the density ratios. The above equation indicates that to minimize the prediction error given a testing set, we should use density ratios as sample weights during the training stage.

The IS framework is very general such that any methods that would produce non-negative weights estimation are applicable for correcting covariate shift. However, very few studies quantify how different combinations of weight estimating methods and risk prediction models could impact model performance. To fill this gap, we compared three commonly used machine learning models combined with different density ratio estimations methods for mitigating covariate shift problems in sepsis detection. We used routinely measured covariates in the EHR, including heart rate, respiratory rate, SpO2, temperature, systolic and diastolic blood pressures to build machine learning models to detect sepsis's onset using the training data. We reported the model performance on the testing data under covariate shift. In particular, the final models output a probability that sepsis is taking place at the current hour. During the training process, we applied two categories of approaches for density ratio estimation: direct approaches and indirect approaches. For direct approaches, we used two methods called Kernel Mean Matching (KMM)\cite{huang2007correcting} and Relative unconstrained Least-Squares Importance Fitting (RuLSIF)\cite{yamada2013relative} to estimate the density ratios for training samples. For indirect approaches, we built probabilistic classifiers to separate training and test samples\cite{bickel2007discriminative} and then derived the density ratios from the classifiers' output using formulas described later. We built sepsis prediction models using three commonly used machine learning models: logistic regression,  random forests, and neural networks. Then, we applied the density ratios during the training process and examined the effect of the density ratios on these machine learning models.

\section*{Methods}
In this study, we used a public available EHR data called eICU Collaborative Research Database (eICU), which is a relational database that contains 200,859 admissions and 139,367 patients between 2014 and 2015 from 335 units at 208 hospitals in the US\cite{pollard2018eicu}. The eICU dataset stores clinical data such as diagnosis, vital signs, lab tests, and drug admissions. The dataset is de-identified to comply with the Health Insurance Portability and Accountability Act (HIPAA). In this study, we only made use of six vital signs, which are frequently available in EHRs. We applied the inclusion criteria to keep patients who are at least $18$ years or older, have at least one measurement for each of the six selected vital signs, and have at least 3 hours of data before the onset. With these criteria applied, there are 1431 patients with sepsis encounters in the cohort in the study. To make sure the case numbers and controls are balanced, we randomly picked 1600 patients without sepsis encounters instead of using all patients without sepsis encounters. To maintain independence among samples, we randomly picked one sepsis onset for each sepsis patient and one random time point for each non-sepsis patient. We used the latest sepsis-3 criteria\cite{singer2016third} to define the onset of sepsis. According to the latest sepsis-3 criteria, sepsis is defined as life-threatening organ dysfunction when organs are injured by a dysregulated response to infection. Organ dysfunction is defined as an increase of the Sequential Organ Failure Assessment (SOFA) Score \cite{vincent1996sofa} by 2 or greater after an infection. The onset of sepsis is defined in our analysis by the first use of antibiotics or microbial sampling. Measured values of patients were divided into 1-hour segments. If a patient had multiple values in one type of measurement within an hour, we used their mean to represent the value of this hour for the measurement. When one measurement was missing at a given hour, it was filled with the patient's last measured value to the missing hour. When the patient did not have any measurement prior to the missing hour, it was filled with the next available measurement.

\begin{figure}[h!]
\centering
\includegraphics[scale=0.4]{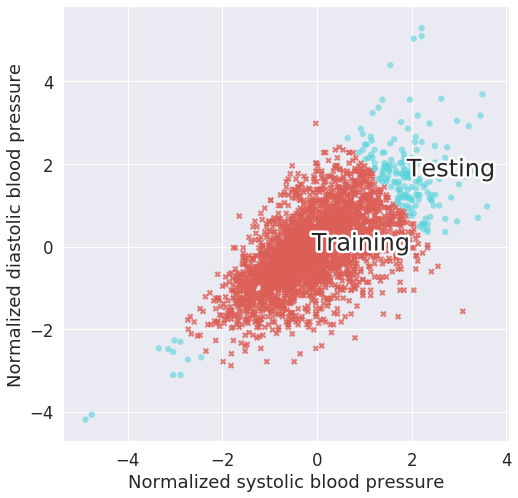}
\caption{Training and testing data sets. Each red point represents a training sample and each blue points represents a testing sample.}
\label{bp_dist}
\end{figure}

Although the eICU dataset contains data from multiple medical centers, there is no information about the patients’ accurate admission times and locations due to the HIPAA regulations. Therefore, we could not develop an approach to obtain the covariate shift based on temporal or geographical differences. Instead, we developed approaches to mimic covariate shift scenarios. In particular, we trained classifiers to detect sepsis when randomly splitting the dataset with 5-fold cross-validation and extracted the mean of covariates importance from the classifiers. Using random forests classifiers, we found that systolic blood pressure and diastolic blood pressure have the highest predictive power among all covariates (see Table \ref{prelm}). We split the dataset into two parts based on cluster membership output from spectral clustering using systolic blood pressure and diastolic blood pressure. Distributions of normalized systolic and diastolic blood pressures of training and test data are displayed in Figure \ref{bp_dist}. As a result, there is a covariate shift between training and testing in terms of systolic blood pressure and diastolic blood pressure. We chose these two variables instead of other variables because if a variable is not critical for predicting the outcome, distribution changes in this variable will not impact risk prediction.\\

\begin{table}[h]
\centering 
\caption{Covariates importance by random forests} \label{prelm}
  \begin{tabular}{|l|l|}
  \hline
     \textbf{Covariate} & \textbf{Importance} \\ \hline
     Systolic blood pressure &  0.217  \\ \hline
     Diastolic blood pressure & 0.232 \\ \hline
     Heart rate &  0.176 \\ \hline
     Respiratory rate & 0.140\\ \hline
     SpO2 & 0.087\\ \hline
     Temperature & 0.147\\ \hline
  \end{tabular}
\end{table}

We applied two categories of approaches to compute the density ratio of training samples: direct estimation and indirect estimation. Kernel Mean Matching (KMM) and Least-Squares Importance Fitting (RuLSIF) are used as direct approaches to estimate density ratios. KMM is proposed by Huang et al.\cite{huang2007correcting} and the main idea of KMM is to match the moment of $r(\mathbf{x})p_{train}(\mathbf{x})$ and $p_{test}(\mathbf{x})$ using kernel functions. The optimization problem of KMM can be solved by quadratic programming. RuLSIF is proposed by Yamada et al.\cite{yamada2013relative} and its main idea is to estimate the relative density ratio by minimizing the squared loss. The optimal solution of RuLSIF can be obtained analytically. 

Besides the direct estimation approaches mentioned above, we also built probabilistic classifiers to calculate the density ratio based on the classifiers’ prediction. The classifiers were trained to predict the probability of a sample coming from the test set given its covariates $\mathbf{x}$, which is denoted as $p$(sample is from test$|\mathbf{x}$). The density ratio can be computed as:
\begin{equation}
    r(\mathbf{x}) = \frac{p(\text{sample is from test set}|\mathbf{x})}{(1 - p(\text{sample is from test set}|\mathbf{x}))}
\end{equation}
Then the ratio was normalized in order to avoid extremely large values. In this study, we chose two commonly used machine learning models, logistic regression with L1 and L2 penalty and random forests, to generate the density ratio indirectly.

Both parametric and non-parametric models were built to detect the onset of sepsis with/without density ratio correction using the training data. We chose random forests\cite{breiman2001random} as a representative of non-parametric models and chose the neural networks and logistic regression with L1 and L2 penalty\cite{zou2005regularization} as representatives of parametric models. For parameters tuning, we performed 5-fold cross-validation to find optimal hyper-parameters with the highest AUROCs. Building these models is essentially a weighted classification with each sample weighted by density ratio estimation. We evaluated the model performance on predicting the onset of sepsis using the testing data. We reported the Area Under the ROC curve (AUROC) as the discrimination metric (a larger value is favored) and the Brier score as the calibration metric (a smaller value is favored), which are two necessary measures to assess the performance of a clinical risk prediction model\cite{alba2017discrimination}. As shown in the literature\cite{davis2017calibration,davis2017calibration2}, discrimination accuracy is critical for evaluating risk prediction model performance, but it does not assess the accuracy of individual risk predictions. Hence, we reported both metrics to provide a more comprehensive assessment. \\

\begin{table}[h]
\centering
\caption{Models' AUROCs under different covariate shift correction methods}\label{auroc}
  \begin{tabular}{|l|c|c|c|}
  \hline
    & \textbf{Logistic Regression}  & \textbf{Random Forests}  &\textbf{Neural Networks} \\ \hline
    \textbf{Training 5-CV}     & 0.801     & 0.824 &  0.816 \\ \hline
    \textbf{Upper bound}        & 0.790      & 0.820 &  0.802 \\ \hline
    \textbf{Without correction} & \shortstack[c]{0.673\\$[0.669, 0.675]$} & \shortstack[c]{0.773\\$ [0.757, 0.791]$} & \shortstack[c]{0.694\\$[0.667, 0.723]$} \\ \hline
    \textbf{With KMM ratio}     & \shortstack[c]{0.684\\$[0.679, 0.692]$} & \shortstack[c]{0.791\\$[0.758, 0.813]$} & \shortstack[c]{0.706\\$[0.676, 0.736]$} \\ \hline
    \textbf{With RuLSIF ratio}  & \shortstack[c]{0.707\\$[0.700,0.710]$} & \shortstack[c]{0.778\\$[0.754, 0.803]$} & \shortstack[c]{0.715\\$[0.686, 0.748]$} \\ \hline
    \textbf{With RF ratio}      & \shortstack[c]{0.676\\$[0.671, 0.681]$} & \shortstack[c]{0.773\\$[0.749, 0.794]$} & \shortstack[c]{0.715\\$[0.681, 0.750]$} \\ \hline
    \textbf{With LR ratio}      & \shortstack[c]{0.680\\$[0.675, 0.684]$} & \shortstack[c]{0.782\\$[0.757, 0.799]$} & \shortstack[c]{0.711\\$[0.683, 0.745]$} \\ \hline
  \end{tabular}
\end{table}

\section*{Results}

Table \ref{auroc} and Table \ref{brier} show the numerical performance of models when predicting the occurrence of sepsis. The first rows are the models' performance on the training data with 5-fold cross validation. We also put the covariates and labels from the testing data together with the training data. The second rows display the models' performance using 5-fold cross validation with all these samples. Therefore, the second rows in the two tables can be considered as the upper bounds of correction methods on a dataset with covariate shift because the covariate shift disappears with such a dataset. The third rows show the models' performance on the testing set without any sample weights applied when models were trained on the training set. The last four rows show the models' performance on the testing set with different correction methods when models were trained on the training set. The logistic regression improved when density ratios were applied as evidenced by a higher AUROC with RuLSIF ($0.707$) when compared to an  AUROC of $0.673$ without any corrections. The improvement upon random forests was limited with respect to the indirect density ratio estimation approaches. In addition, AUROC improvement was also observed in the neural networks with the density ratio corrections, but the Brier scores were not improved with the incorporation of KMM and RuLSIF ratios, which indicates that the improvement in AUROC is not necessarily tied together with the improvement in the Brier scores.

\begin{table}[h]
\centering
\caption{Models' Brier scores under different covariate shift correction methods}\label{brier}
  \begin{tabular}{|l|c|c|c|}
  \hline
    & \textbf{Logistic Regression}  & \textbf{Random Forests}  &\textbf{Neural Networks} \\ \hline
    \textbf{Training 5-CV}     & 0.181      & 0.173 &  0.174 \\ \hline
    \textbf{Upper bound}        & 0.186      & 0.174 &  0.180 \\ \hline
    \textbf{Without correction} & \shortstack[c]{0.250\\$[0.248, 0.253]$} & \shortstack[c]{0.185\\$ [0.181, 0.189]$} & \shortstack[c]{0.210\\$[0.196, 0.236]$} \\ \hline
    \textbf{With KMM ratio}     & \shortstack[c]{0.211\\$[0.204, 0.220]$} & \shortstack[c]{0.182\\$ [0.173, 0.191]$} & \shortstack[c]{0.236\\$ [0.208, 0.259]$} \\ \hline
    \textbf{With RuLSIF ratio}  & \shortstack[c]{0.224\\$[0.220, 0.231]$} & \shortstack[c]{0.184\\$ [0.177, 0.191]$} & \shortstack[c]{0.234\\$[0.209, 0.257]$} \\ \hline
    \textbf{With RF ratio}      & \shortstack[c]{0.233\\$ [0.230, 0.238]$} & \shortstack[c]{0.185\\$ [0.180, 0.191]$} & \shortstack[c]{0.207\\$ [0.185, 0.233]$} \\ \hline
    \textbf{With LR ratio}      & \shortstack[c]{0.235\\$ [0.232, 0.239]$} & \shortstack[c]{0.182\\$ [0.176, 0.188]$} & \shortstack[c]{0.209\\$[0.187, 0.233]$} \\ \hline
  \end{tabular}
\end{table}

\begin{figure}[h!]
\centering
\includegraphics[scale=0.30]{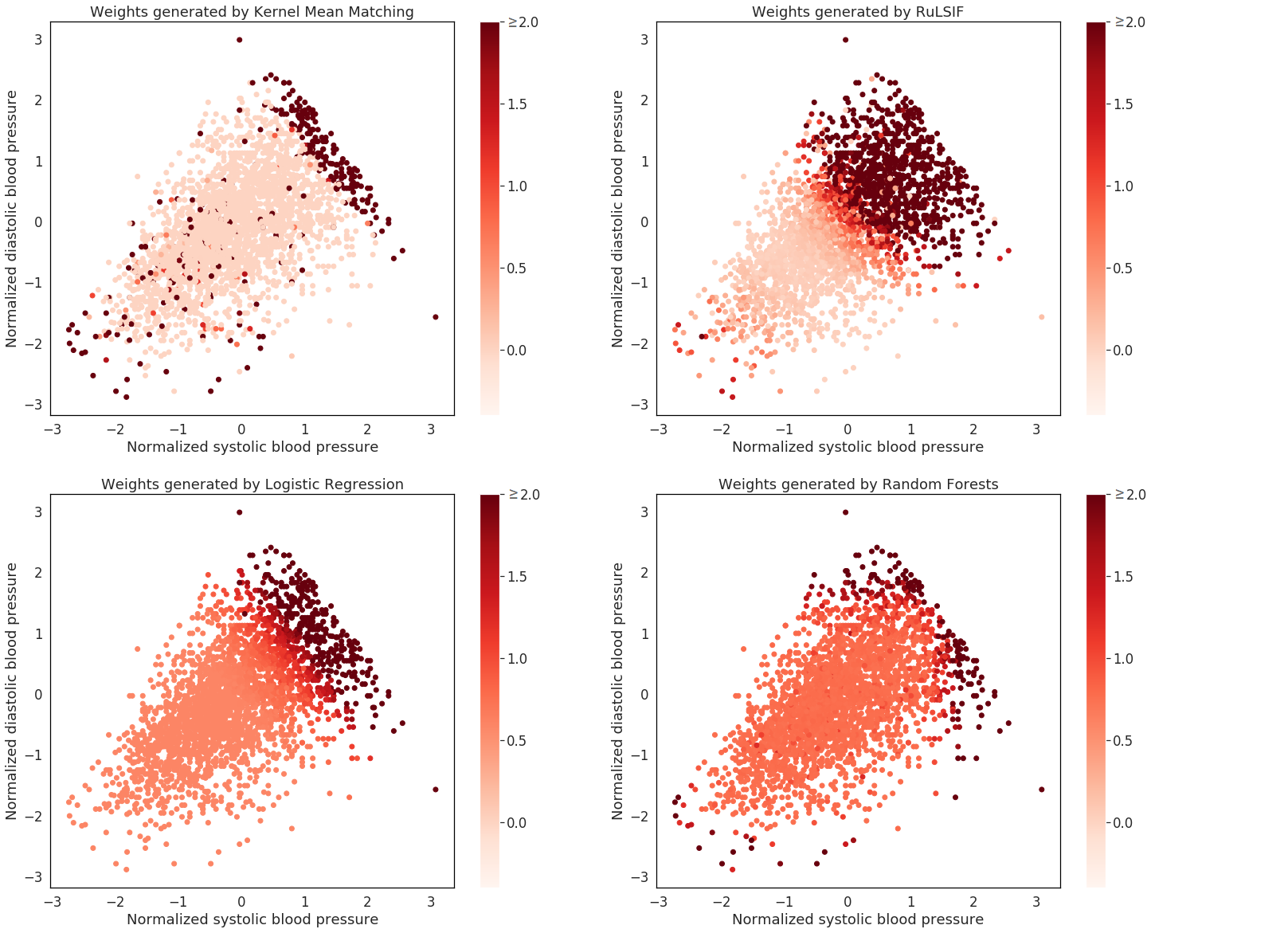}
\caption{Visualization of training sample weights}
\label{drs}
\end{figure}

Figure \ref{drs} shows the derived training samples density ratios. Sample points with darker colors are assigned with more substantial weights, whereas sample points with lighter colors are assigned with smaller weights. Since the testing data are mainly located near the edges of the plot, the results from Figure \ref{drs} indicate that samples which are similar to the test set are assigned heavier weights by both direct and indirect covariate shift corrections. 

\begin{figure}[h!]
\centering
\includegraphics[scale=0.30]{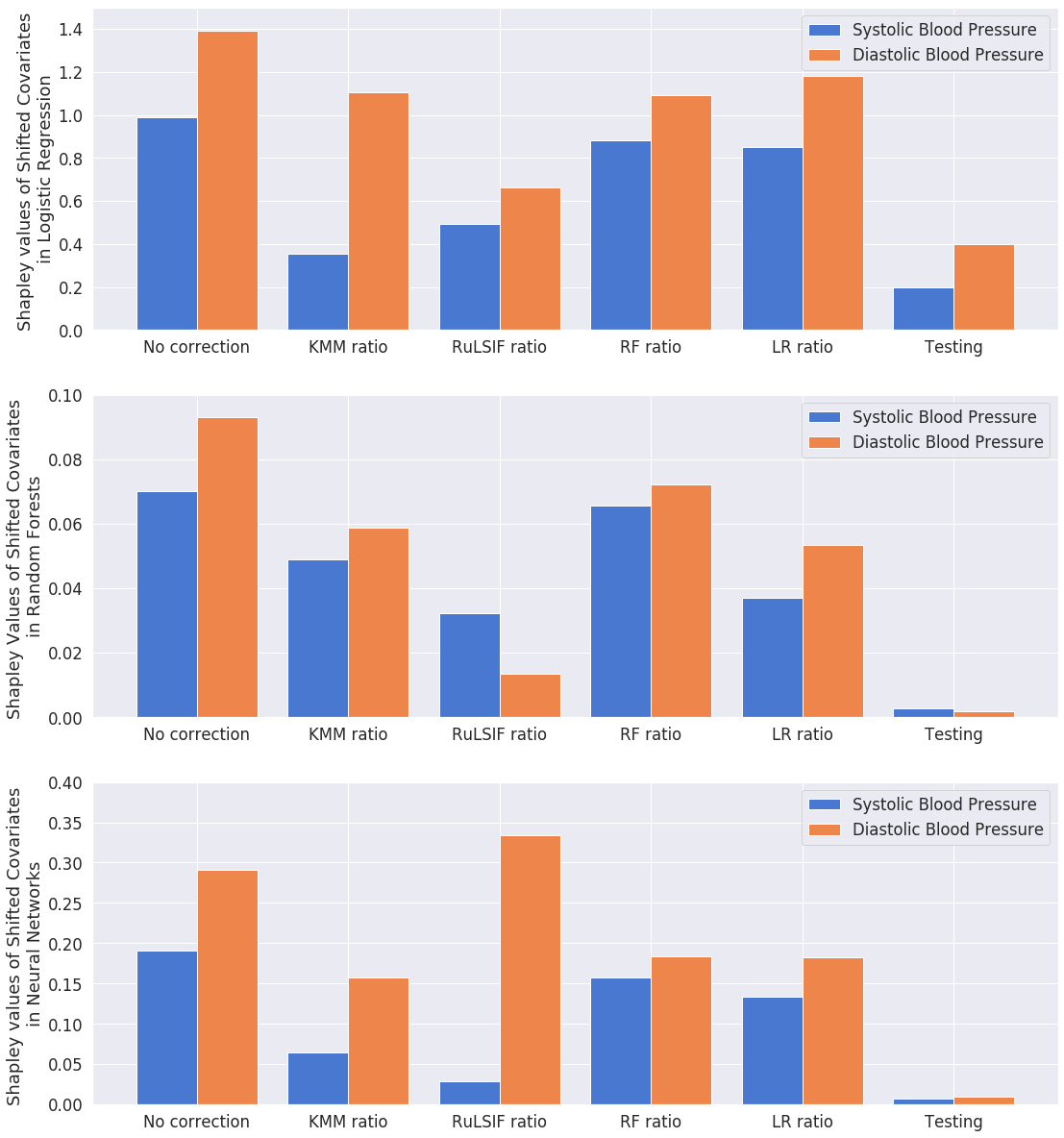}
\caption{Change of feature importance (Shapley values) and coefficients before and after the corrections. The Shapley values are computed through game theories and can explain the covariates’ contribution to a model’s predictions. A covariate with a large Shapley value indicates that it significantly influences the model given other covariates }
\label{shap}
\end{figure}

To track the change of shifted covariates' impact on the models, we obtained the Shapley values of systolic blood pressures and diastolic blood pressures. The Shapley values are computed through game theories and can explain the covariates' contribution to a model's predictions\cite{kalai1987weighted,lundberg2017unified}. A covariate with a large Shapley value indicates that it significantly influences the model given other covariates. Figure \ref{shap} shows the Shapley values of systolic blood pressures and diastolic blood pressures before and after the corrections, as well as on the models which were learned using the testing set (denoted as "Testing" group). For the logistic regression models, the Shapley values are close to the model trained on the testing data, especially with the RuLSIF ratios. For the random forests and neural networks, the values are far from the "Testing" group, even though they are different from the model without any correction.



\section*{Discussion}

In this work, we applied covariate shift corrections to sepsis detection task for ICU patients. We found that when covariate shift takes place, assigning different weights to training samples based on their similarity to testing samples can improve the performance of machine learning models. Compared with some previous studies\cite{mitra2018sepsis,mao2018multicentre}, we adopted much fewer measurements as covariates/features and achieved an AUROC close to $0.80$ with covariates correction. Our work is potentially impactful on clinical practice, especially when we want to transit a risk prediction model trained using one healthcare system to another healthcare system where the population may have inherently different characteristics. Our results indicate that applying covariate shift corrections are likely to make the model more generalizable. Although our paper focused on detecting sepsis onset, the strategy of correcting the covariate shift is applicable to detecting and predicting other clinical risks without much modification.

Compared with the random forests model (a complex nonlinear model), we found that logistic regression is more sensitive to the density ratio corrections and achieved a bigger improvement in discrimination and calibration. We believe that the random forests model is more complex and less likely to suffer from the model misspecification. Random forests could achieve decent performance even before the correction (AUROC larger than $0.75$ and Brier scores smaller than $0.20$ for random forests before corrections compared with AUROC smaller than $0.70$ and Brier scores larger than $0.25$ for logistic regression). In contrast, the logistic regression is accurate only when the linear functions of vital signs can well approximate the logit of being sepsis cases. Hence, the density ratio can alleviate the model misspecification problem by encouraging the model to perform well on the testing data instead of training data. Similar to the random forests, Brier scores of the neural network models did not improve much after density ratio correction. One possible explanation is that the neural networks with nonlinear activation layers are flexible nonlinear models, making them suffer less from misspecification than logistic regression. These results indicate that parametric/simple models are more likely to benefit from the covariate shift corrections than complex models.

The Brier scores of the neural networks show that density ratio correction does not necessarily improve both discrimination and calibration of the machine learning models. One explanation is that the hyper-parameters are optimized through cross-validation using AUROC (a discrimination metric); hence the corrections may have less impact on calibration metrics such as the Brier score. On the other hand, the variable importance of covariates (in particular, the shifted covariates) could also change after applying density ratio correction. Such a change of variable importance is accompanied by the improvement of the models’ performance. For example, the AUROC of the logistic regression model increased from $0.673$ to $0.707$ with the RuLSIF ratio applied, and the corresponding Shapley values of the shifted covariates are relatively close to the values of the model trained on the testing set. In contrast, the performance of the random forests and neural networks model have minimal improvement after correction as the Shapley values of the density ratio corrected models are very different from the model trained on the testing data.

\begin{table}[h]
\centering 
\caption{Model performance (AUROC) with prior shift} \label{prior}
  \begin{tabular}{|l|l|l|}
  \hline
      & \shortstack[l]{First Test set \\with 45.9\% positive} & \shortstack[l]{Second Test set \\with 55.6\% positive}  \\ \hline
     Random Forests & 0.836 & 0.836  \\ \hline
     Logistic Regression & 0.809  & 0.806    \\ \hline
     Neural Networks & 0.814  & 0.820    \\ \hline
  \end{tabular}
\end{table}

In this paper, we focused on the effect of the covariate shift. However, other types of data shift can also happen in practice. Based on Equation (2), we have shown that in theory, the density ratio correction is useful for covariate shift, while density ratio correction may not be necessary/useful for prior shift and concept shift. In particular, the prior shift can lead to case/control imbalance, which could have a mild impact on model generalizability. Resampling based approaches that are useful for handling imbalanced classification could potentially remedy the problem caused by prior shift. On the other hand, density ratio correction is not helpful for concept shift at all. We have conducted simulations to support these two conclusions as follows. To create a prior shift, we used $1958$ of the original training samples, where 46.9\% are positive, to train classifiers then tested the classifiers on two testing set: one is the $451$ samples from the original training samples where 45.9\% are sepsis cases, the other is the remaining $450$ samples from the original training samples where 55.6\% are sepsis cases. The AUROC performance displayed in Table \ref{prior} demonstrates that the prior shift does not substantially affect model performance for our setting. To create a concept shift, we used 1958 samples as the training set and two sets with precisely the same sample covariates as testing. Then, we replaced the outcome labels of the second testing set with labels generated from a model that is independent of the sample covariates. Such a setting guarantees that both testing sets have no covariate shift and only the second testing set has a concept shift. We tested the covariate shift correction, and the results in Table \ref{concept} show that the density ratio cannot mitigate the performance drop caused by the concept shift. How to remedy the problem caused by concept shift is an open question and interesting future direction. 

\begin{table}[h]
\centering 
\caption{Model performance (AUROC) with concept shift} \label{concept}
  \begin{tabular}{|l|l|l|l|}
  \hline
    & Logistic Regression  & Random Forests & Neural Networks  \\ \hline
    Training 5-CV      & 0.801 & 0.824 & 0.808 \\ \hline
    Upper bound        & 0.720 & 0.787 & 0.745 \\ \hline
    Wihtout correction & 0.394 & 0.381 & 0.393 \\ \hline
    With KMM ratio     & 0.364 & 0.383 & 0.376 \\ \hline
    With RuLSIF ratio  & 0.373 & 0.391 & 0.334 \\ \hline
    With RF ratio      & 0.379 & 0.438 & 0.334 \\ \hline
    With LR ratio      & 0.385 & 0.383 & 0.386 \\ \hline
  \end{tabular}
\end{table}

Over the past two decades, most sepsis-related studies have used the Sepsis-2 criteria as the gold standard for sample labeling cases of sepsis \cite{mitra2018sepsis,mao2018multicentre,lyra2019early}. The Sepsis-2 criteria uses vital signs such as heart rates and body temperature to define the onset of sepsis\cite{kaukonen2015systemic}. This definition may lead to data leakage if we use the same measurements in both outcome labeling and feature construction. Therefore, we used the latest Sepsis-3 criteria as the gold standard in this project. Besides, previous studies\cite{mitra2018sepsis,mao2018multicentre,lyra2019early} only used AUC to measure model performance. On the one hand, AUC has desirable properties such as scale-invariance and being concordance based; on the other hand, AUC is a discrimination metric that cannot reflect calibration. For this reason, we also reported Brier score (a calibration metric) together with the AUROC to better characterize the models’ performance improvement.

\section*{Conclusion}
We built parametric and non-parametric models to detect the onset of sepsis for ICU patients. Direct and indirect approaches are applied to compute density ratios for covariate shift correction.  We found that both parametric and non-parametric models benefit from the density ratios but the simpler parametric models such as the logistic regression are more sensitive to the covariate shift corrections.

\section*{Acknowledgement}
This study was partially funded through a Fall Research Competition grant from OVCRGE at University of Wisconsin-Madison and through Patient‐Centered Outcomes Research Institute (PCORI) Awards (ME-2018C2-13180). The views in this paper are solely the responsibility of the authors and do not necessarily represent the views of the PCORI, its Board of Governors or Methodology Committee.

\makeatletter
\renewcommand{\@biblabel}[1]{\hfill #1.}
\makeatother

\bibliographystyle{vancouver}
\bibliography{amia}  

\begin{thebibliography}{10}

\bibitem{sakr2018sepsis}
Sakr Y, Jaschinski U, Wittebole X, Szakmany T, Lipman J, {\~N}amendys-Silva SA,
  et~al.
\newblock Sepsis in intensive care unit patients: worldwide data from the
  intensive care over nations audit.
\newblock In: Open forum infectious diseases. vol.~5. Oxford University Press
  US; 2018. p. ofy313.

\bibitem{rhee2016diagnosing}
Rhee C, Kadri SS, Danner RL, Suffredini AF, Massaro AF, Kitch BT, et~al.
\newblock Diagnosing sepsis is subjective and highly variable: a survey of
  intensivists using case vignettes.
\newblock Critical Care. 2016;20(1):89.

\bibitem{singer2016third}
Singer M, Deutschman CS, Seymour CW, Shankar-Hari M, Annane D, Bauer M, et~al.
\newblock The third international consensus definitions for sepsis and septic
  shock (Sepsis-3).
\newblock Jama. 2016;315(8):801--810.

\bibitem{mitra2018sepsis}
Mitra A, Ashraf K.
\newblock Sepsis prediction and vital signs ranking in intensive care unit
  patients.
\newblock arXiv preprint arXiv:181206686. 2018;.

\bibitem{mao2018multicentre}
Mao Q, Jay M, Hoffman JL, Calvert J, Barton C, Shimabukuro D, et~al.
\newblock Multicentre validation of a sepsis prediction algorithm using only
  vital sign data in the emergency department, general ward and ICU.
\newblock BMJ open. 2018;8(1).

\bibitem{pollard2018eicu}
Pollard TJ, Johnson AE, Raffa JD, Celi LA, Mark RG, Badawi O.
\newblock The eICU Collaborative Research Database, a freely available
  multi-center database for critical care research.
\newblock Scientific data. 2018;5:180178.

\bibitem{kumar2006duration}
Kumar A, Roberts D, Wood KE, Light B, Parrillo JE, Sharma S, et~al.
\newblock Duration of hypotension before initiation of effective antimicrobial
  therapy is the critical determinant of survival in human septic shock.
\newblock Critical care medicine. 2006;34(6):1589--1596.

\bibitem{subbe2006validation}
Subbe C, Slater A, Menon D, Gemmell L.
\newblock Validation of physiological scoring systems in the accident and
  emergency department.
\newblock Emergency Medicine Journal. 2006;23(11):841--845.

\bibitem{rangel1995natural}
Rangel-Frausto MS, Pittet D, Costigan M, Hwang T, Davis CS, Wenzel RP.
\newblock The natural history of the systemic inflammatory response syndrome
  (SIRS): a prospective study.
\newblock Jama. 1995;273(2):117--123.

\bibitem{vincent1996sofa}
Vincent JL, Moreno R, Takala J, Willatts S, De~Mendon{\c{c}}a A, Bruining H,
  et~al.. The SOFA (Sepsis-related Organ Failure Assessment) score to describe
  organ dysfunction/failure.
\newblock Springer-Verlag; 1996.

\bibitem{lyra2019early}
Lyra S, Leonhardt S, Antink CH.
\newblock Early Prediction of Sepsis Using Random Forest Classification for
  Imbalanced Clinical Data.
\newblock In: 2019 Computing in Cardiology (CinC). IEEE; 2019. p. 1--4.

\bibitem{dexter2020generalization}
Dexter GP, Grannis SJ, Dixon BE, Kasthurirathne SN.
\newblock Generalization of Machine Learning Approaches to Identify Notifiable
  Conditions from a Statewide Health Information Exchange.
\newblock AMIA Summits on Translational Science Proceedings. 2020;2020:152.

\bibitem{quionero2009dataset}
Quionero-Candela J, Sugiyama M, Schwaighofer A, Lawrence ND.
\newblock Dataset shift in machine learning.
\newblock The MIT Press; 2009.

\bibitem{moreno2012unifying}
Moreno-Torres JG, Raeder T, Alaiz-Rodr{\'\i}Guez R, Chawla NV, Herrera F.
\newblock A unifying view on dataset shift in classification.
\newblock Pattern recognition. 2012;45(1):521--530.

\bibitem{hwang2019development}
Hwang EJ, Park S, Jin KN, Im~Kim J, Choi SY, Lee JH, et~al.
\newblock Development and validation of a deep learning--based automated
  detection algorithm for major thoracic diseases on chest radiographs.
\newblock JAMA network open. 2019;2(3):e191095--e191095.

\bibitem{nestor2019feature}
Nestor B, McDermott M, Boag W, Berner G, Naumann T, Hughes MC, et~al.
\newblock Feature robustness in non-stationary health records: caveats to
  deployable model performance in common clinical machine learning tasks.
\newblock arXiv preprint arXiv:190800690. 2019;.

\bibitem{curth2019transferring}
Curth A, Thoral P, van~den Wildenberg W, Bijlstra P, de~Bruin D, Elbers P,
  et~al.
\newblock Transferring clinical prediction models across hospitals and
  electronic health record systems.
\newblock In: Joint European Conference on Machine Learning and Knowledge
  Discovery in Databases. Springer; 2019. p. 605--621.

\bibitem{sugiyama2007covariate}
Sugiyama M, Krauledat M, M{\~A}{\v{z}}ller KR.
\newblock Covariate shift adaptation by importance weighted cross validation.
\newblock Journal of Machine Learning Research. 2007;8(May):985--1005.

\bibitem{huang2007correcting}
Huang J, Gretton A, Borgwardt K, Sch{\"o}lkopf B, Smola AJ.
\newblock Correcting sample selection bias by unlabeled data.
\newblock In: Advances in neural information processing systems; 2007. p.
  601--608.

\bibitem{yamada2013relative}
Yamada M, Suzuki T, Kanamori T, Hachiya H, Sugiyama M.
\newblock Relative density-ratio estimation for robust distribution comparison.
\newblock Neural computation. 2013;25(5):1324--1370.

\bibitem{bickel2007discriminative}
Bickel S, Br{\"u}ckner M, Scheffer T.
\newblock Discriminative learning for differing training and test
  distributions.
\newblock In: Proceedings of the 24th international conference on Machine
  learning; 2007. p. 81--88.

\bibitem{breiman2001random}
Breiman L.
\newblock Random forests.
\newblock Machine learning. 2001;45(1):5--32.

\bibitem{zou2005regularization}
Zou H, Hastie T.
\newblock Regularization and variable selection via the elastic net.
\newblock Journal of the royal statistical society: series B (statistical
  methodology). 2005;67(2):301--320.

\bibitem{alba2017discrimination}
Alba AC, Agoritsas T, Walsh M, Hanna S, Iorio A, Devereaux P, et~al.
\newblock Discrimination and calibration of clinical prediction models:
  users’ guides to the medical literature.
\newblock Jama. 2017;318(14):1377--1384.

\bibitem{davis2017calibration}
Davis SE, Lasko TA, Chen G, Matheny ME.
\newblock Calibration drift among regression and machine learning models for
  hospital mortality.
\newblock In: AMIA Annual Symposium Proceedings. vol. 2017. American Medical
  Informatics Association; 2017. p. 625.

\bibitem{davis2017calibration2}
Davis SE, Lasko TA, Chen G, Siew ED, Matheny ME.
\newblock Calibration drift in regression and machine learning models for acute
  kidney injury.
\newblock Journal of the American Medical Informatics Association.
  2017;24(6):1052--1061.

\bibitem{kalai1987weighted}
Kalai E, Samet D.
\newblock On weighted Shapley values.
\newblock International journal of game theory. 1987;16(3):205--222.

\bibitem{lundberg2017unified}
Lundberg SM, Lee SI.
\newblock A unified approach to interpreting model predictions.
\newblock In: Advances in neural information processing systems; 2017. p.
  4765--4774.

\bibitem{kaukonen2015systemic}
Kaukonen KM, Bailey M, Pilcher D, Cooper DJ, Bellomo R.
\newblock Systemic inflammatory response syndrome criteria in defining severe
  sepsis.
\newblock New England Journal of Medicine. 2015;372(17):1629--1638.

\end{thebibliography}

\end{document}